\documentclass[useAMS,usenatbib]{mn2e}

\usepackage{graphicx}
\usepackage{subfigure}

\title[The ADC for the LAMOST]{The atmospheric dispersion corrector for the Large Sky Area Multi--object Fibre Spectroscopic Telescope (LAMOST)}
\author[Ding-qiang Su, Peng Jia and Genrong Liu]{Ding-qiang Su$^{1,2,3}$\thanks{E-mail:
dqsu@nju.edu.cn}, Peng Jia$^{1,2,3}$ and Genrong Liu$^{3}$\\
$^{1}$Department of Astronomy, Nanjing University, 22 Hankou Road, Nanjing 210093, China\\
$^{2}$Key Laboratory of Modern Astronomy and Astrophysics (Nanjing University), Ministry of Education, Nanjing 210093, China\\
$^{3}$National Astronomical Observatories / Nanjing Institute of Astronomical Optics \& Technology (NIAOT), Chinese Academy of Science,\\  188 Bancang Street, Nanjing 210042, China}
\begin{document}

\date{}

\pagerange{\pageref{firstpage}--14} \pubyear{2011}

\maketitle

\label{firstpage}

\begin{abstract}
The Large Sky Area Multi--object Fibre Spectroscopic Telescope (LAMOST) is the largest (aperture 4 $\mathrm{m}$) wide field of view (FOV) telescope and is equipped with the largest amount (4000) of optical fibres in the world. For the LAMOST North and the LAMOST South the FOV are 5 $^\circ$ and 3.5 $^\circ$, the linear diameters are 1.75 $\mathrm{m}$ and 1.22 $\mathrm{m}$, respectively. A new kind of atmospheric dispersion corrector (ADC) is put forward and designed for LAMOST. It is a segmented lens which consists of many lens--prism strips. Although it is very big, its thickness is only 12 $\mathrm{mm}$. Thus the difficulty of obtaining big optical glass is avoided, and the aberration caused by the ADC is small. Moving this segmented lens along the optical axis, the different dispersions can be obtained. The effects of ADC's slits on the diffraction energy distribution and on the obstruction of light are discussed. The aberration caused by ADC is calculated and discussed. All these results are acceptable. Such an ADC could also be used for other optical fibre spectroscopic telescopes, especially those which a have very large FOV.
\end{abstract}


\section{Introduction}

Large Sky Area Multi--object Fibre Spectroscopic Telescope (LAMOST) is a new type telescope \citep{Wang96,Cui2000,Su2004}. The main parameters of LAMOST are the following: clear aperture 4 $\mathrm{m}$ (average), f--ratio 5, and field of view (FOV) 5 $^\circ$. The linear diameter of FOV is 1.75 $ \mathrm{m}$. 4000 optical fibres \citep{xing1998}, which introduce the light of different celestial objects to 16 spectrographs \citep{Zhu2006}, are put on such a big focal surface. The dedication ceremony of LAMOST was held on 2008 October 16. We call it the LAMOST North. At present, another telescope of this kind, the LAMOST South designed to survey the southern sky, is under consideration by China and other countries, with a view to international cooperation \citep{Cui20102}. The clear aperture and f--ratio of the LAMOST South are still 4 $\mathrm{m}$ and $5$. But its FOV will be reduced to 3.5 $^\circ$. The linear diameter of FOV of this telescope is 1.22 $\mathrm{m}$. Many years ago some atmospheric dispersion correctors for small FOV had been designed and used. Since 1980's some atmospheric dispersion correctors for larger FOV have been designed \citep{Epps1984,Su1986,SuLiang1986,wynne1986,willstrop1987,liang1988,Su1988,Bingham1988,wynne1988, wangsu1990}. In these correctors there is a pair of prisms or lens--prisms (lensms), each of which is a cemented lens with a tilted cemented surface and consists of two different glasses. Rotating these two lens--prisms, we could obtain different dispersions. Using this method in LAMOST would require very big lens--prisms, and it is difficult to obtain optical glass of such large size and to support the big lens--prisms. Liu \& Yuan have designed several kinds of small corrector each for an optical fibre \citep{LiuYuan2009}. But it is difficult to install and move them for 4000 optical fibres. In this paper, as an example a detailed design and discussion of this ADC are given for the LAMOST South. It is a big but thin segmented lens which consists of many lens--prism strips. Thus the difficulty of obtaining such big optical glass is avoided, and since it is thin, the introduced aberration is small. Moving this segmented lens along the optical axis, we could obtain different dispersions. The entire design of this ADC could be used for the LAMOST North too---we only need to extend its diameter to about 1.78 $\mathrm{m}$. In this ADC the clear aperture of a portion of celestial objects is divided into two parts, {\rmfamily i.e.}, a slit is added in it. Here the word ``silt'' means silt plus chamfer at the edge of each lens--prism strip and they are covered with black paint, {\rmfamily i.e.}, in this paper ``slit'' means a light-obstructing black belt with a width of 1 $\mathrm{mm}$. In this case the diffraction spot is enlarged. But in LAMOST, as the following discussion will show, it is not serious and it is acceptable. Apart from this, there is some loss from light obstruction by the silt. For different celestial objects, the loss is different in the whole FOV, and also different when ADC is at different positions. For these two reasons mentioned above, such a segmented ADC is not suitable for the diffraction--limited high resolution telescope and photometry. Such a segmented ADC is mainly used for the optical fibre spectroscopic telescope, especially if this telescope has a very large FOV.

\section{A brief introduction of LAMOST}

\begin{figure*}
    \centering
    \includegraphics[scale=0.4]{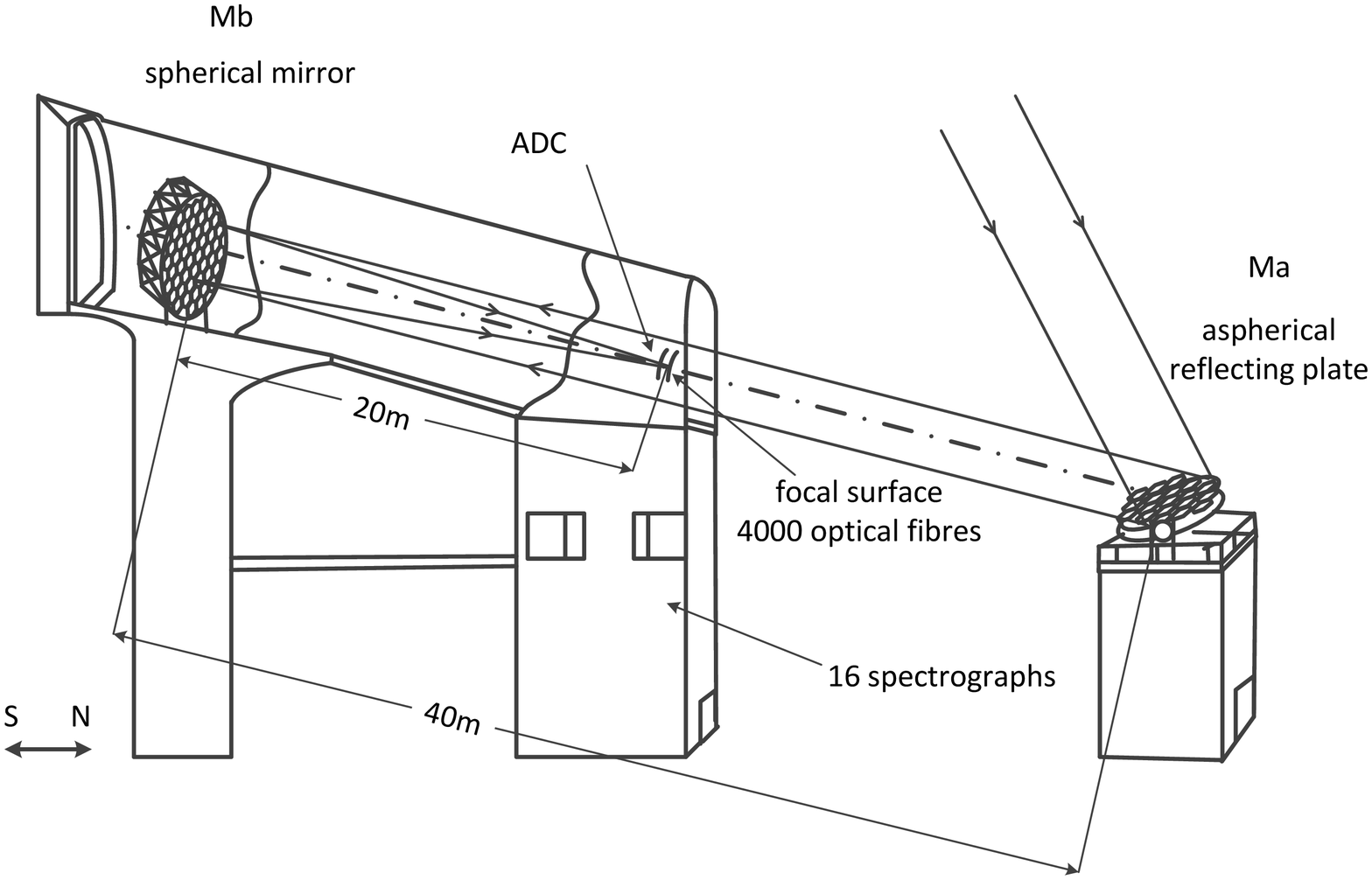}
    \caption{The LAMOST and the position of the ADC in LAMOST}
    \label{fig1}
    \end{figure*}

LAMOST is shown in Figure \ref{fig1}. It lies on the ground along the South--North direction. Mb is a spherical mirror. Ma is an aspherical reflecting plate, which corrects the spherical aberration of Mb and reflects the light of celestial objects to the Mb. Ma and Mb consists of 24 and 37 hexagonal mirrors respectively and each mirror has diagonal 1.1 $\mathrm{m}$. In a segmented mirror telescope PSF is enlarged, especially when a segmented ADC is added. At a good astronomical site, if we do not use adaptive optics, atmospheric seeing only allows a telescope with an aperture under 30 $\mathrm{cm}$ to obtain the diffraction--limited image. Because adaptive optics cannot be used for a large FOV telescope, therefore, in this case for a segmented mirror telescope a sub-area of diameter 50--60 $\mathrm{cm}$ is enough. The other reason that LAMOST North uses two segmented mirrors is to save the cost. By the way, it is possible that LAMOST South will use one segmented mirror consisting of 1.1 $\mathrm{m}$ hexagonal sub--mirrors. On the other hand, in many optical fibre spectroscopic telescopes the diameter of optical fibre is more than 1.5 $\mathrm{arc sec}$. Compared with the seeing and diameter of the optical fibre, the effect of enlarged PSF in LAMOST including segmented ADC is minor and acceptable. During the observation, for a particular observation direction (a particular sky area), LAMOST is a reflecting Schmidt telescope. But when observing different directions ({\rmfamily i.e.} different celestial objects or same celestial objects in different times), LAMOST should be different reflecting Schmidt telescopes. This means the shape of Ma should be different for correcting the spherical aberration. Traditionally, such an optical system can not be realized. Active optics is a key technology for correcting the telescope errors--gravitation deformation and thermal deformation. This technology was developed mainly by Wilson and Lemaitre {\rmfamily et.al.} \citep{Wilson1999,Lam2009} for the thin mirror, and by Nelson and Mast et.al. \citep{Nelson1980,MaNel1980,MaNel1982} for the segmented mirror. Chinese experts have creatively applied active optics technology to Ma to make such a telescope possible and developed the thin--mirror and segmented--mirror combined active optics \citep{Su1996,Su1998,Cui2000,Su2004,cui2008}. LAMOST is a wide FOV telescope with the largest aperture and has the strongest fibre spectroscopic obtaining capability in the world. The observing sky area of LAMOST is $-10^\circ\leqslant\delta\leqslant+90^\circ$. The celestial objects are observed for an average of 1.5 hours before and after they pass through the meridian. During observation only the mounting of Ma does the tracking and the focal surface does the rotation. LAMOST North was set up in Xing Long Station (latitude 40.4 $^\circ$ N, height above sea level 900 $ \mathrm{m}$), National Astronomical Observatories, Chinese Academy of Sciences. The biggest zenith distance (with the celestial object in the meridian, it is the same in the following text) is 50.4 $^\circ$. For waveband $380$ -- $1000 \mathrm{nm}$ the atmospheric dispersion is 2.2 $ \mathrm{arc sec}$. At Xing Long Station FWHM of seeing is about 2 $\mathrm{arc sec}$. The biggest spread of aberration is 1.84 $\mathrm{arc sec}$. The diameter of optical fibre adopted is 3.3 $\mathrm{arc sec}$. In this situation, although it is better to correct the atmospheric dispersion, leaving it uncorrected will not present serious problem. Through the development of the LAMOST North, Chinese experts found that the size of each optical fibre positioner could be significantly reduced, so in the LAMOST South the FOV will be reduced to 3.5 $^\circ$, the linear diameter of it is 1.22 $ \mathrm{m}$, and 6000 optical fibres could be put on this focal surface. The observing sky area of the LAMOST South is $0^\circ\geqslant\delta\geqslant-90^\circ$. The LAMOST South may be installed on the the NOAO Las Campanas Observatory or the ESO Paranal Observatory. NOAO Las Campanas Observatory is situated at latitude 29.9 $^\circ$ S, altitude 2400 $\mathrm{m}$ above the sea level and the biggest zenith distance observed is 60.1 $^\circ$. ESO Paranal Observatory is situated at latitude 24.6 $^\circ$ S, altitude 2635 $\mathrm{m}$ above the sea level and the biggest zenith distance observed is 65.4 $^\circ$. For waveband $380$ -- 1000 $\mathrm{nm}$ the corresponding atmospheric dispersion at these two observatories are
2.75 $\mathrm{arc sec}$ and 3.35 $\mathrm{arc sec}$ respectively. Since image quality (due to the reduction of FOV to 3.5 $^\circ$) and two observatories'seeing are better than the LAMOST North, the diameter of optical fibre used will be 1.6 $\mathrm{arc sec}$. In the LAMOST South the atmospheric dispersion should be corrected. Since at ESO Paranal Observatory the biggest atmospheric dispersion is bigger, it is chosen as an example in this paper.

\section{The structure of this ADC}

\begin{figure*}
    \centering
    \includegraphics[scale=0.4]{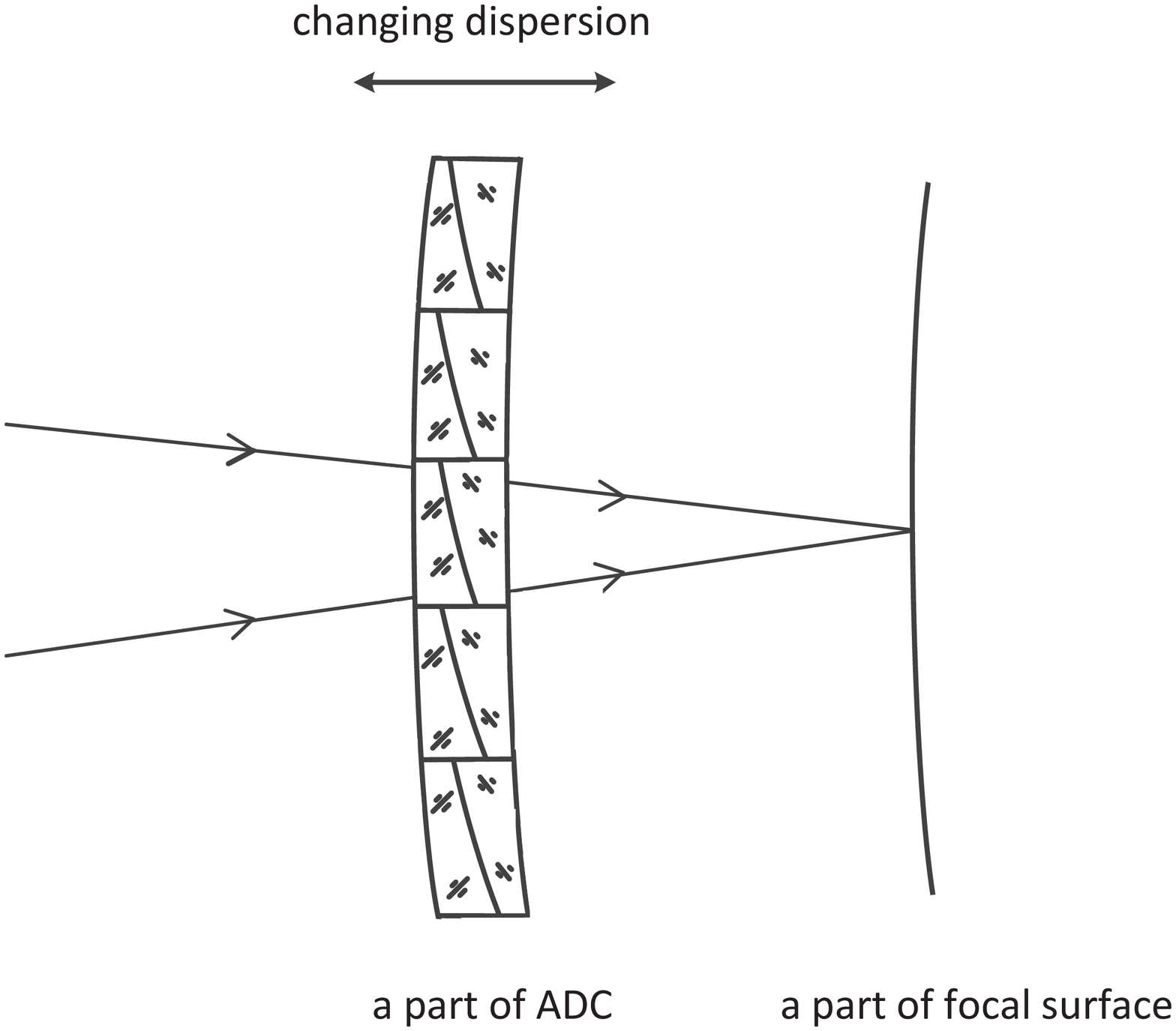}
    \caption{A sectional drawing of the ADC and focal surface. In this figure the radii of ADC and focal surface have been reduced (i.e., more bended), and the thickness of the ADC has been enlarged.}
    \label{fig2}
    \end{figure*}

\begin{figure*}
    \subfigure[]
    {\includegraphics[scale=0.25]{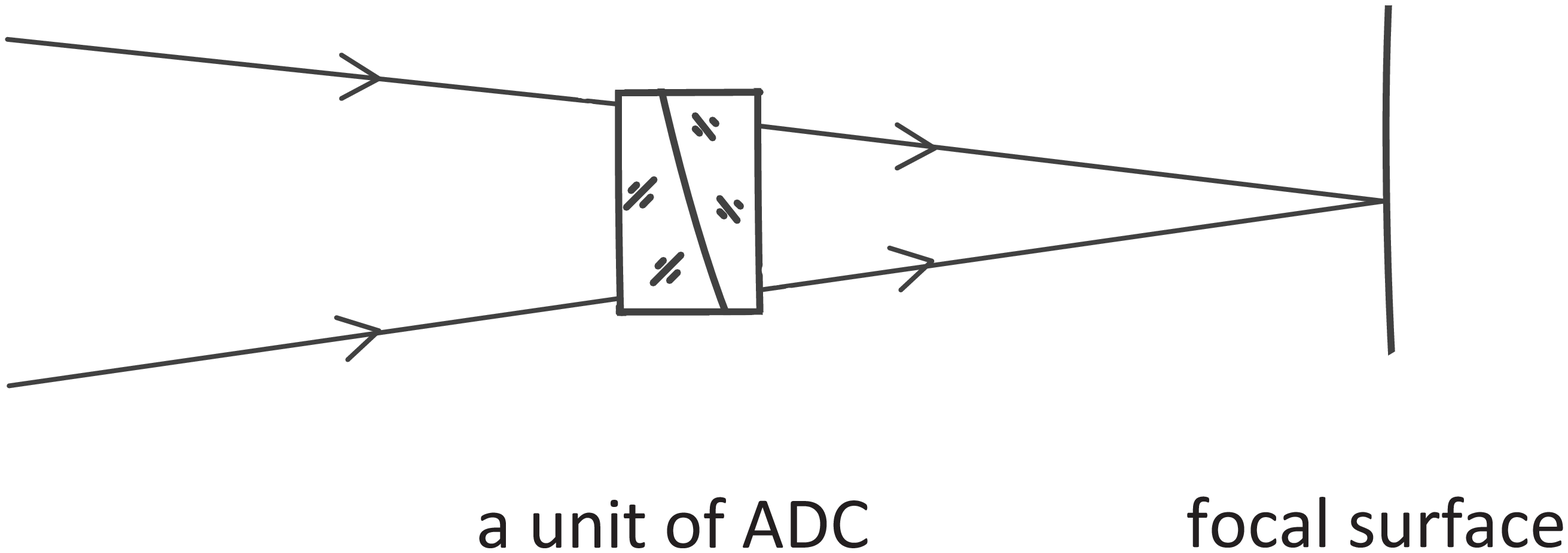}}\label{fig3a}
    \subfigure[]
    {\includegraphics[scale=0.25]{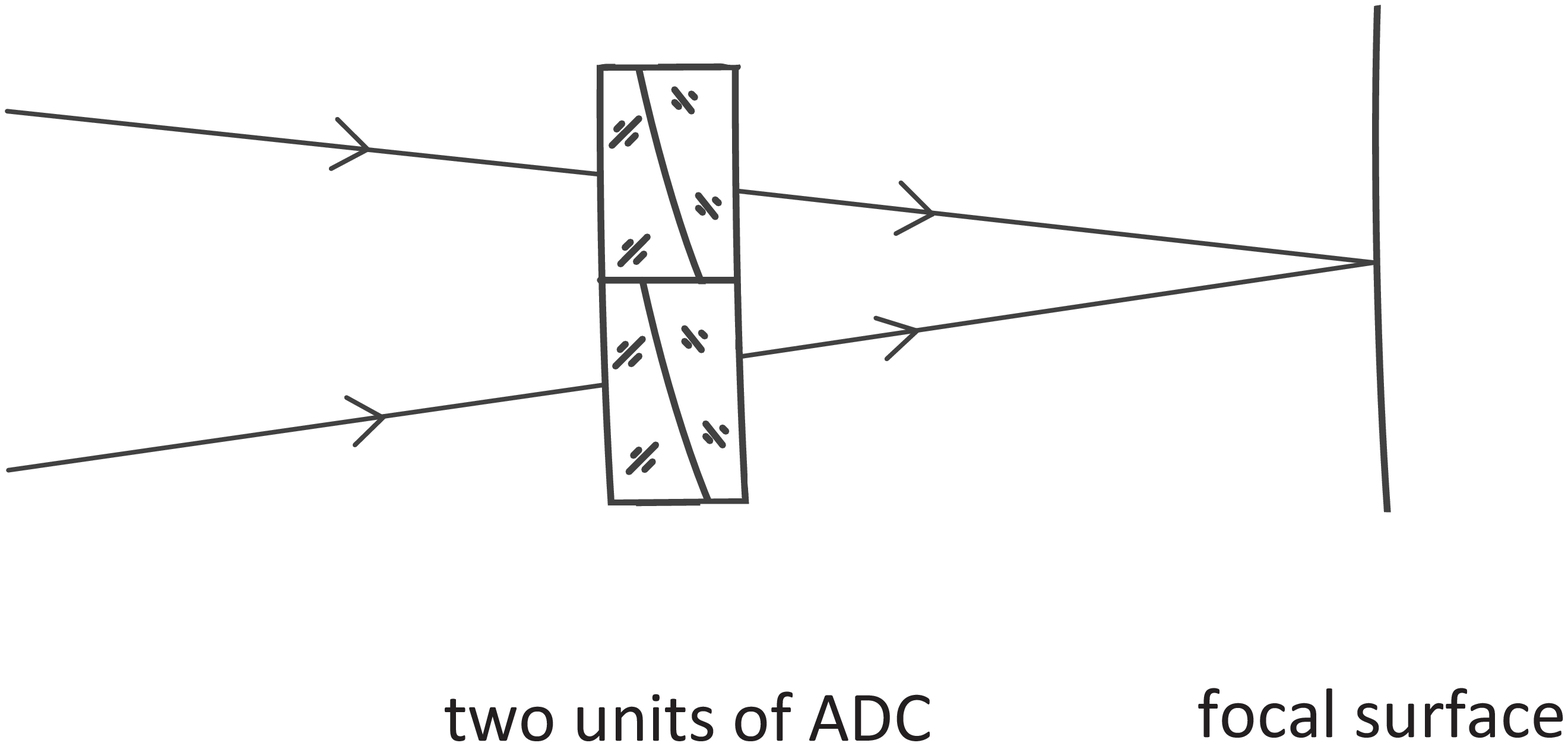}}\label{fig3b}
    \caption{The figure (a) shows the light beam area of the object is at the middle of a lens--prism strip and the figure (b) shows a slit of two lens--prism strips is at the centre of the light beam area of the object.}
    \label{fig3}
    \end{figure*}

The layout and specification of this ADC are shown in Figure \ref{fig2},\ref{fig3} and Table \ref{t1}. Each lens--prism strip consists of Schott glass PSK3 and LLF1.
The refractive index of PSK3 and LLF1 are given in Table \ref{t2}.
In $\lambda$ $=$ 441.8 $\mathrm{nm}$ the refractive indices of the two kinds of glass are the same.
In LAMOST the Mb and the focal surface are concentric.
The radius of focal surface is 20 $\mathrm{m}$.
When ADC is at the farthest position, we take its two outside surfaces and the focal surface to be concentric, {\rmfamily i.e.} the radius of the outside surface equals 20 $\mathrm{m}$ + 250 $\mathrm{mm}$. We require that in this position for waveband $380$ -- 1000 $\mathrm{nm}$ this ADC can produce dispersion of 3.35 $\mathrm{arc sec}$.
From it the tilt angle of lens--prism strip can be obtained which is 6.89 $^\circ$.
We set the width of each lens--prism strip to be 50 $\mathrm{mm}$ which equals to the light beam diameter of each celestial object on this ADC.
Thus in this position only one slit is added for each object.
Since the thickness difference of one lens of lens--prism strip is 6.04 $\mathrm{mm}$, we take the total thickness of ADC is 12 $\mathrm{mm}$. Since the f--ratio of LAMOST is 5, the diameter of ADC should equal the linear FOV diameter plus 50 $\mathrm{mm}$ $(250/5= 50)$, {\rmfamily i.e.}, 1.27 $ \mathrm{m}$ for the LAMOST South. So the maximum length of the strip is also 1.27 $ \mathrm{m}$ for the LAMOST South.
A similar result for the LAMOST North is 1.78 $\mathrm{m}$. If one feels that some lens--prism strips are too long in the LAMOST North, those strips longer than 1 $\mathrm{m}$ could be divided into two parts. In this case, only about $1/20$ celestial objects will meet two slits when ADC is at the farthest position. Although such an ADC is very big, it is very thin and only includes one segmented lens. As it is moved along the optical axis its dispersion is changed. Thus the atmospheric dispersion for $z<65.4$ $^\circ$ celestial objects can be compensated. The dispersion produced by this ADC is direct proportion to the distance from it to focal surface. And in these situations only one slit is met for a part of celestial objects. When the atmospheric dispersion is small enough which needs not to be compensated, this ADC can be moved out easily. With regard to the manufacturing of the ADC, we plan to glue these lens--prism stripes together (with  removable glue) to form a disk, then grind and polish it. We have already had some experience with such a method. Since the maximum light beam of each celestial object on ADC is only 50 $\mathrm{mm}$, the figure tolerance of lens--prism strip is loose. As a whole this ADC's tolerance of position, tip and tilts are loose. All lens--prism strips are fixed on the edge of the frame. So long as each lens-prism strip is well fixed, its tip and tilts will be small. Liquid glue may be used for the cemented surface to reduce the reflecting loss. Nevertheless, moderate difficulties do exist for the manufacturing and mounting of the ADC, thus it is still necessary to conduct more research and testing in this respect.

\section{Some discussions on special topics}

\subsection{The effect of ADC's slit on diffraction energy distribution}\label{slitdiff}
   In LAMOST both Ma and Mb consist of hexagonal mirrors each with a diagonal of 1.1 $\mathrm{m}$. The surface area of such a hexagonal mirror equals to a circular area with a diameter of 1 $\mathrm{m}$. For different celestial objects these sub--mirrors of Ma and Mb are covered by each other in the clear aperture with different states. Since in LAMOST all sub--mirrors only co--focus, the light from different sub--areas is non--coherent and these shapes of sub--areas divided by edges of hexagonal mirrors are complex. First we discuss the case when both Ma and Mb are approximately perpendicular to the optical axis, {\rmfamily i.e.}, the angle between the light of the celestial object and the optical axis pointing to Mb is small. It can be found that each hexagonal sub--mirror is mainly divided into 3 to 4 sub--areas. If we assume 4 sub--areas, an important conclusion can be obtained: the average surface area of sub--area equals a circular area with a diameter of 0.5 $\mathrm{m}$. As a rough estimate we think that the diffraction energy distribution of LAMOST in this case is like a circular hole with a diameter of 0.5 $\mathrm{m}$, {\rmfamily i.e.}, 84 $\mathrm{per cent}$ of the light energy spreads in about 0.5 $\mathrm{arc sec}$ area (Airy disk). Xu made a detailed calculation for four special situations, and a similar conclusion was obtained \citep{Xu97}. Since LAMOST's clear aperture is 4 $\mathrm{m}$, its surface area is 64 times of a 0.5 $\mathrm{m}$ circular area. We could think that about 64 sub--areas are included in LAMOST clear aperture. For a particular celestial object in the worst situation its clear aperture is divided by an ADC's slit along the diameter direction, thus about eight sub--areas are divided. As an average result the diffraction energy of the eight sub--areas will distribute in two times its original length in perpendicular direction to the slit, {\rmfamily i.e.}, the 84 $\mathrm{per cent}$ diffraction energy will spread in an area 0.5 $\mathrm{arc sec}$ wide and 1 $\mathrm{arc sec}$ long, {\rmfamily i.e.}, half of the energy will disperse beyond the 0.5 $ \mathrm{arc sec}$ area. Thus in the 0.5 $\mathrm{arc sec}$ area the light energy will reduce by $4/64$, {\rmfamily i.e.}, about 6 $\mathrm{per cent}$. The energy loss is small and it still distributes in 1 $\mathrm{arc sec}$ area. Given that the diameter of optical fibre is 1.6 $\mathrm{arc sec}$, the energy loss is acceptable. Secondly, we discuss the situation where the celestial objects observed are near the celestial pole. In this case, in a plane perpendicular to the optical axis the projective width of sub--mirrors of Ma will reduce to about 1/2, but the projective width of sub--mirrors of Mb is unchangeable.  Considering these two factors and using a method of analysis similar to the above, we obtain the following conclusion: in this case in LAMOST 84 $\mathrm{per cent}$ of the light energy spreads in an area about 0.5 $\mathrm{arc sec}$ wide and 0.75 $\mathrm{arc sec}$ long and in this area the light energy will reduce about 4.5 $\mathrm{per cent}$ due to an ADC's slit. The energy loss is small and it distributes in a 1.5 $\mathrm{arc sec}$ area. Given that the diameter of optical fibre is 1.6 $\mathrm{arc sec}$, the energy loss is acceptable. In LAMOST South either Ma or Mb may adopt a non--segmented (monolithic) mirror. It is easy to find that if either Ma or Mb is a non--segmented mirror or consists of larger sub--mirrors, with such a segmented ADC the total diffraction energy distribution will be more concentrated than in the above situation.

  \subsection{The probability of objects meet slit}\label{slit}
         In the largest zenith distance $\mathrm{z}$ = 65.4 $^\circ$, each object will meet a slit of strips. When $\mathrm{z}<$65.4 $^\circ$, {\rmfamily i.e.}, the distance between ADC and focal surface is less than 250 $\mathrm{mm}$, only a part of objects meet a slit of lens--prism strips. For example, if the distance between ADC and focal surface is 100 $\mathrm{mm}$ (according to $\mathrm{z}$ = 39.6 $^\circ$, see section \ref{cal} and Table \ref{tab3}) only $2/5$ objects meet a slit and if the distance between ADC and focal surface is 50 $\mathrm{mm}$ (according to $\mathrm{z} = $20.6 $^\circ$), only $1/5$ objects meet a slit.

 \subsection{The light obstructed by slits of ADC}\label{lilos}
According to a technical requirement, the edge of strips of ADC should be chamfered to a projective width of 0.5 $\mathrm{mm}$. In order to reduce scattered light, all chamfers and slits of ADC should be covered with black paint. Thus all slits will become black belts with a width of 1 $\mathrm{mm}$ between two strips. These slits will obstruct light. The width of each strip is 50 $\mathrm{mm}$. The approximate average obstructed ratio equals 1/50 = 2 $\mathrm{per cent}$. The thickness of ADC is 12 $\mathrm{mm}$. The f--ratio of LAMOST is 5 and the maximum inclination angle of ray is 1/10 to ADC's surface. Considering these situations, we obtain that about an average of 2.5 $\mathrm{per cent}$ of light will be obstructed. This is the average light loss. For a specific celestial object, this loss may be zero when the object does not meet a slit, and it may be several times the average loss when ADC is near the focal surface and the object meets a slit. Since the loss of light obstruction by a slit is uneven, such a segmented ADC is not suitable for photometry.

 \subsection{The ADC's orientation}\label{ori}
        Both the atmospheric dispersion and ADC's dispersion are vectors. The compensating error equals ADC dispersion vector plus the atmospheric dispersion vector. For compensation not only the amounts of the two vectors should be equal but also their directions should be opposite to each other. In a telescope the ADC should be rotated to make its dispersion direction opposite to the atmospheric dispersion direction. From spherical astronomy the ADC's orientation formula can be derived easily. The tolerance of ADC's orientation angular is loose.

\subsection{The aberrations}\label{abe}
      First, we ignore the tilt of cemented surface of ADC. Since the radii of ADC is very large, it can be considered as a parallel glass plate. From the third--order aberration formula the least circle of spherical aberration is only 0.01 $\mathrm{arc sec}$ and it could be corrected by active optics. From the chromatic aberration formula the spread circle of it is 0.17 $\mathrm{arc sec}$ at the extreme wavelengths 380 $ \mathrm{nm}$ and 1000 $\mathrm{nm}$. For a glass parallel plate the spherical aberration and chromatic aberration are unchanged when it moved along optical axis. When the tilt of cemented surface of ADC is considered the aberration mainly coma will be added. It increases with the increasing difference of the refractive index of two glasses, {\rmfamily i.e.}, it is maximum in two extreme wavelengths. And this aberration is direct proportion to the distance from ADC to the focal surface. Here we do not use formula to calculate it. In next section by using \textsc{Zemax} software all aberrations, indicated by spot diagrams, will be given including this aberration.

   \subsection{The compensation error of atmospheric dispersion}\label{compen}
      Due to the difference between the atmosphere and the glass dispersion, the compensation error is existent. It is just like as secondary spectrum in an achromatic optical system. Apparently it is direct proportion to the amount of the compensated atmospheric dispersion.

\section{The calculation and following discussion}\label{cal}
\begin{figure*}
    \includegraphics[scale=0.3,angle=90]{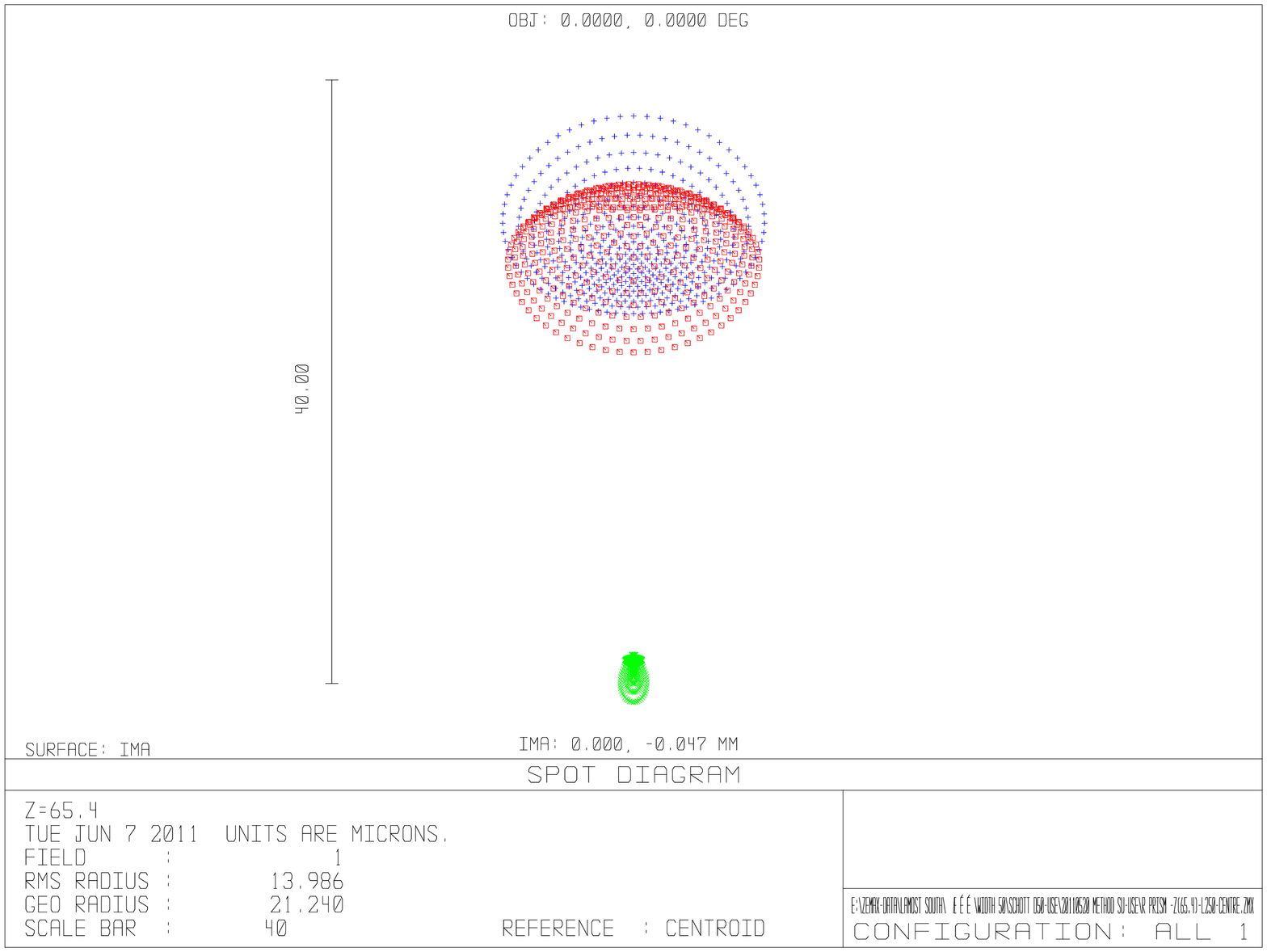}

    \includegraphics[scale=0.3,angle=90]{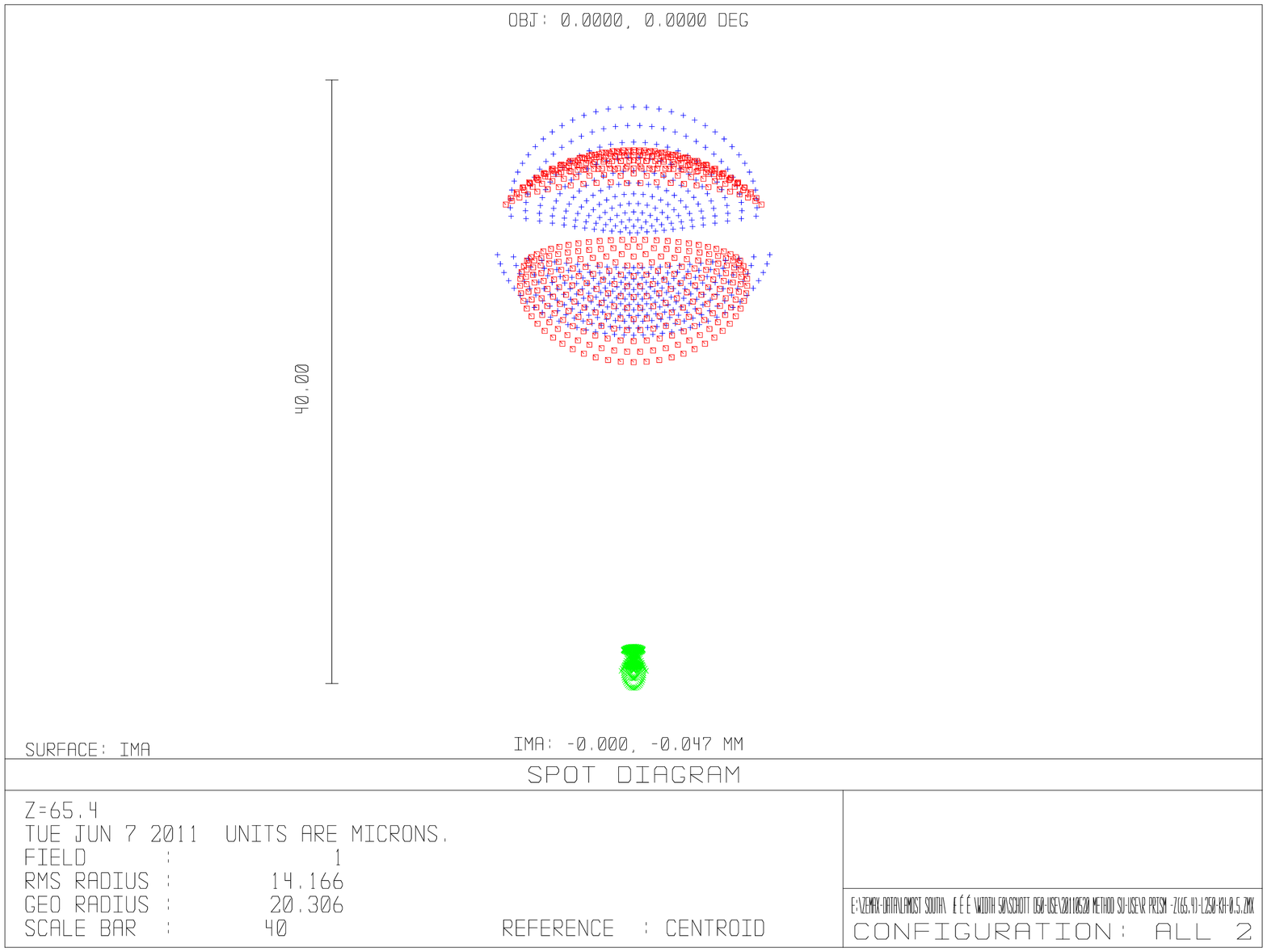}
 \caption{The spot diagram when the distance from the first surface of the ADC to the focal surface is 250 $\mathrm{mm}$.
  The upper figure shows the spot diagram when the light beam area of the object is at the middle of a lens--prism strip and the lower figure shows when a slit of two lens--prism strips is at the centre of the light beam area of the object.
  The length of the scale line (the line beside the spot diagram) represents 40 $\mu\mathrm{m}$ (0.413 $\mathrm{arc sec}$). The different colour of the diagrams in the picture represent different wavelength:
  blue, $\lambda$ 380 $\mathrm{nm}$, the spot diagram in the top of the upper figure;
  green, $\lambda$ 500 $\mathrm{nm}$, the spot diagram in the lowermost of the upper figure;
  red, $\lambda$ 1000 $\mathrm{nm}$, the spot diagram in the middle of the upper figure. }
  \label{fig4}
  \end{figure*}

\begin{figure*}
  \includegraphics[scale=0.3,angle=90]{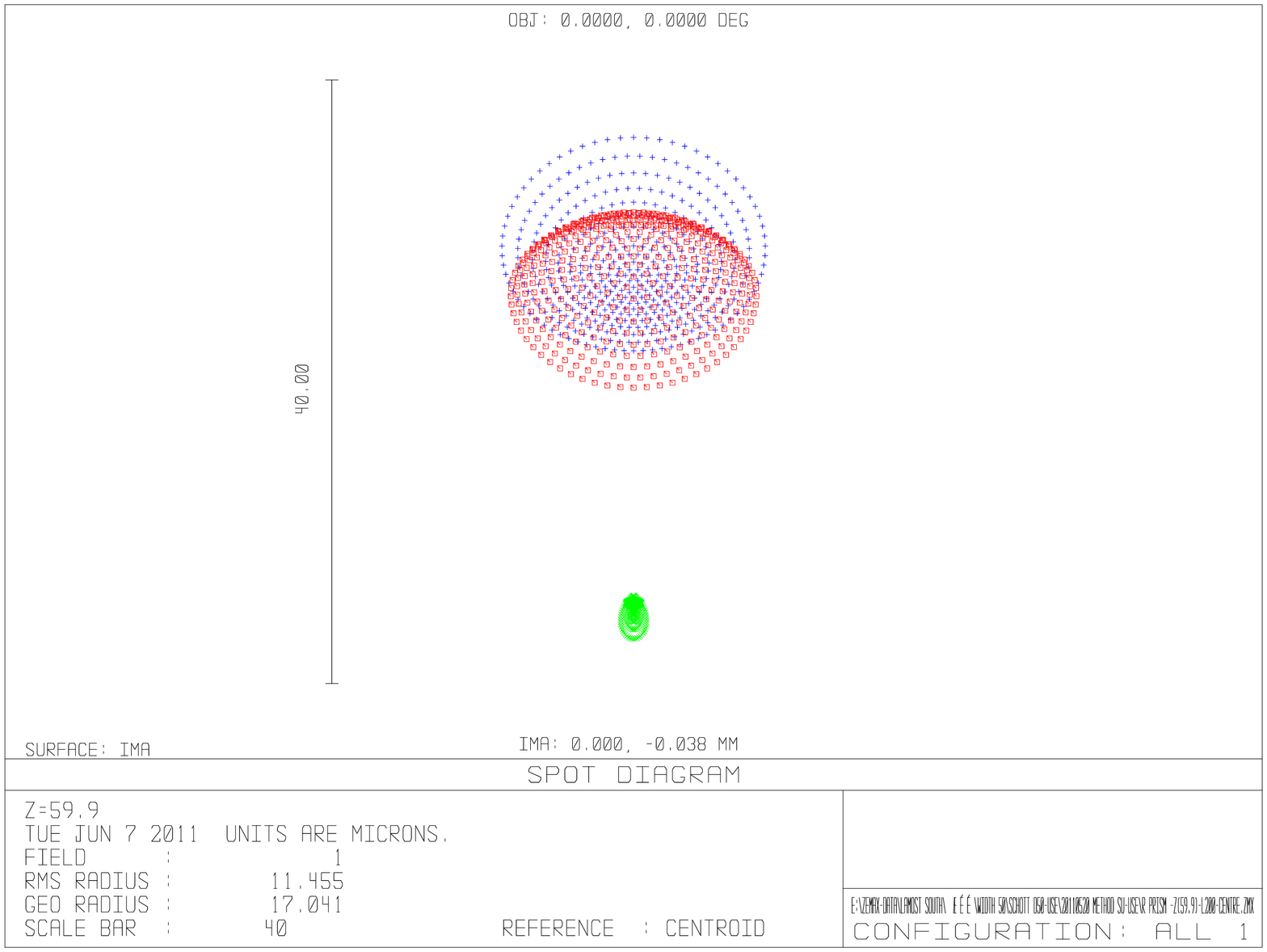}
  \includegraphics[scale=0.3,angle=90]{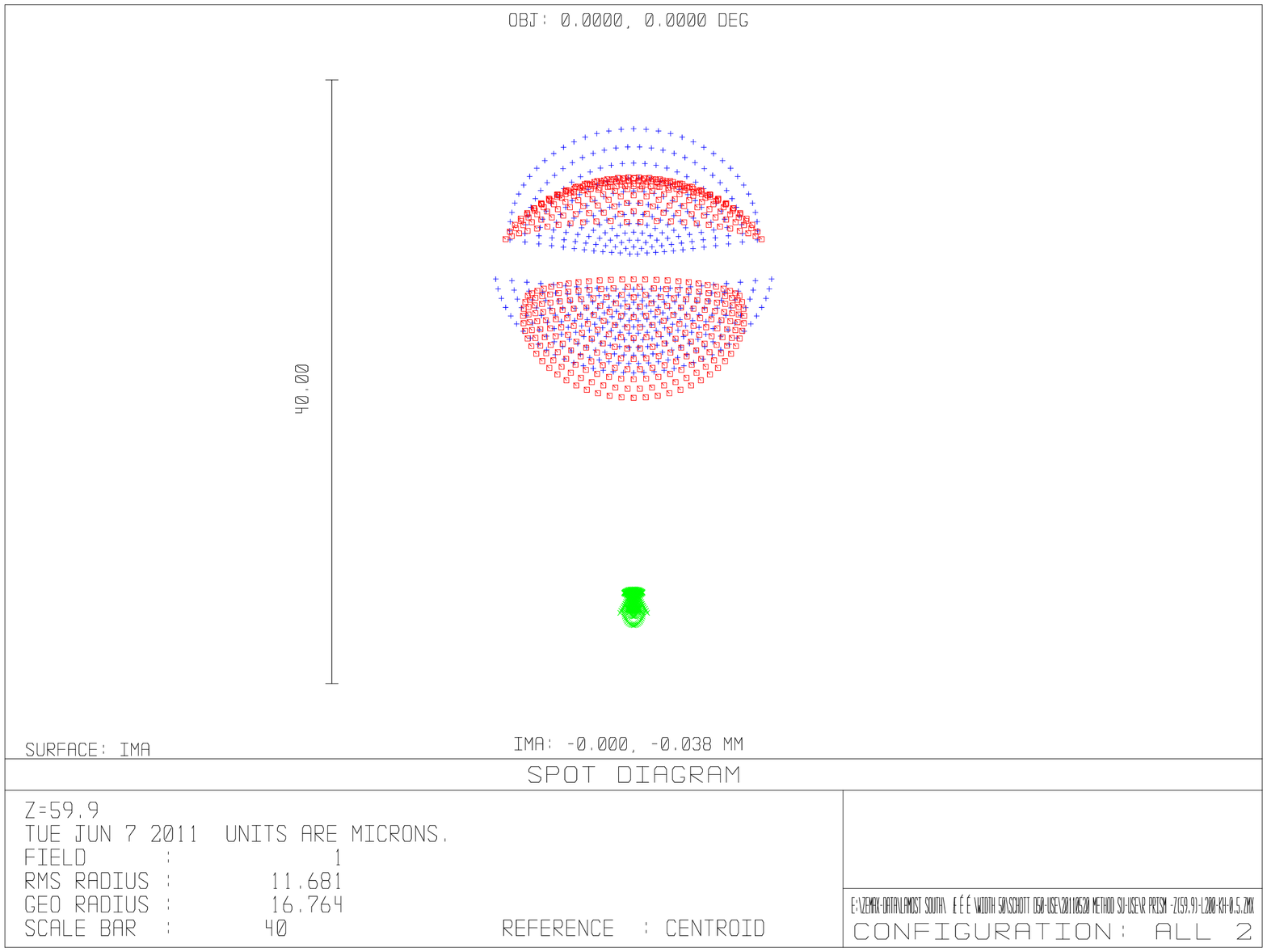}
  \caption{The spot diagram when the distance from the first surface of ADC to the focal surface is 200 $\mathrm{mm}$. The rest of explanation is the same as in Figure \ref{fig4}.}
  \label{fig5}
  \end{figure*}

\begin{figure*}
  \includegraphics[scale=0.3,angle=90]{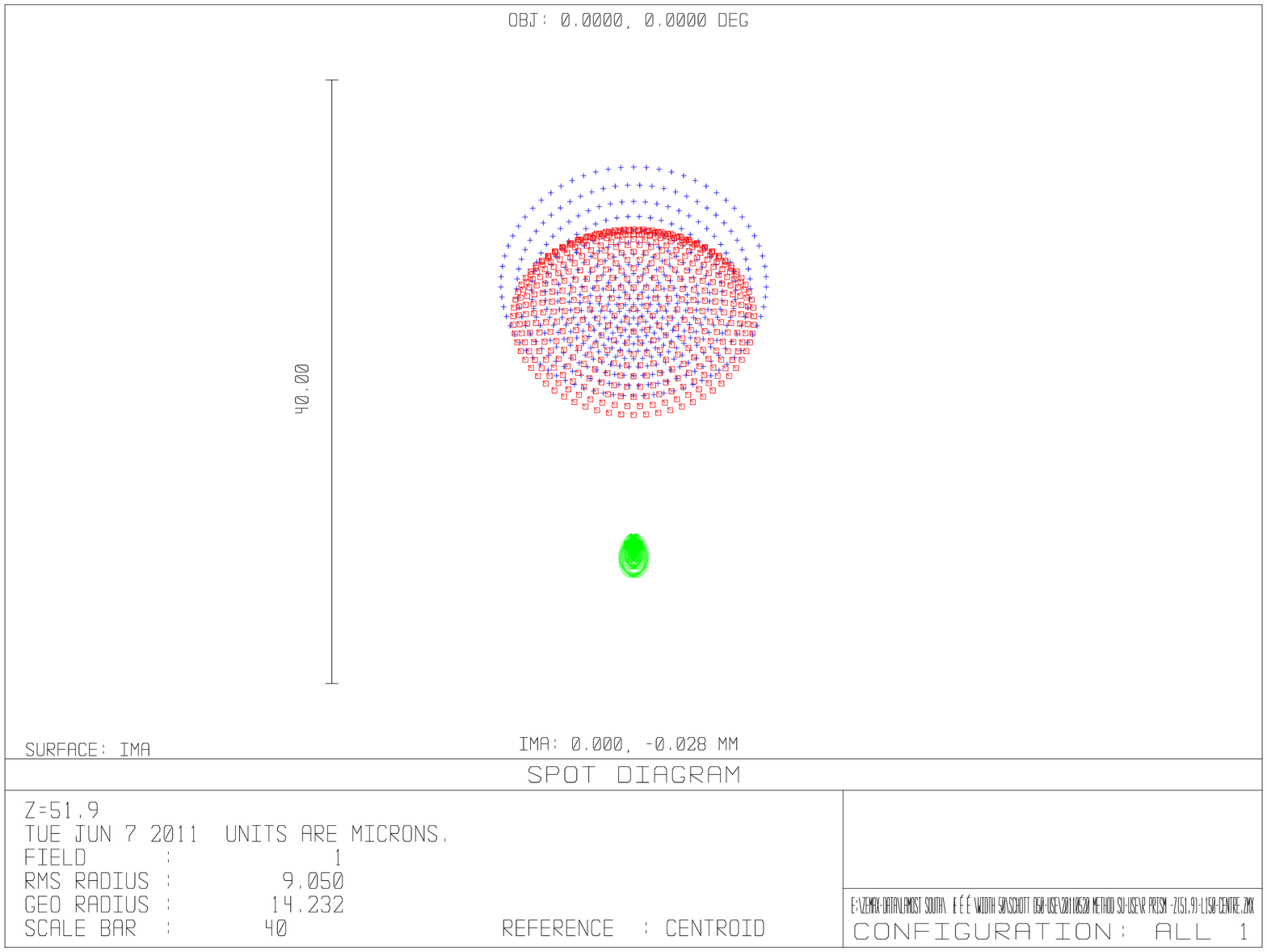}
  \includegraphics[scale=0.3,angle=90]{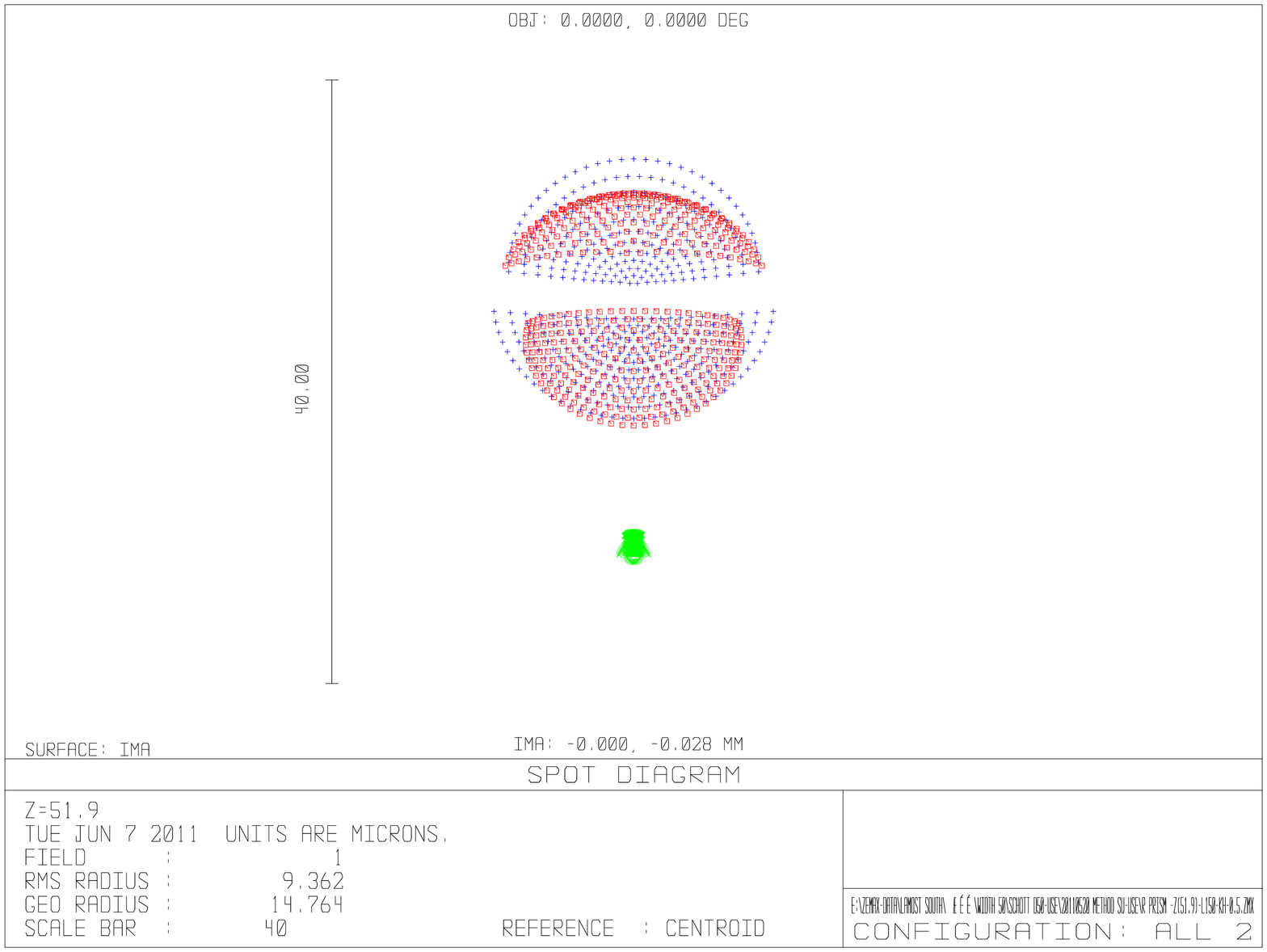}
  \caption{The spot diagram when the distance from the first surface of ADC to the focal surface is 150 $\mathrm{mm}$. The rest of explanation is the same as in Figure \ref{fig4}.}
  \label{fig6}
  \end{figure*}

\begin{figure*}
  \includegraphics[scale=0.3,angle=90]{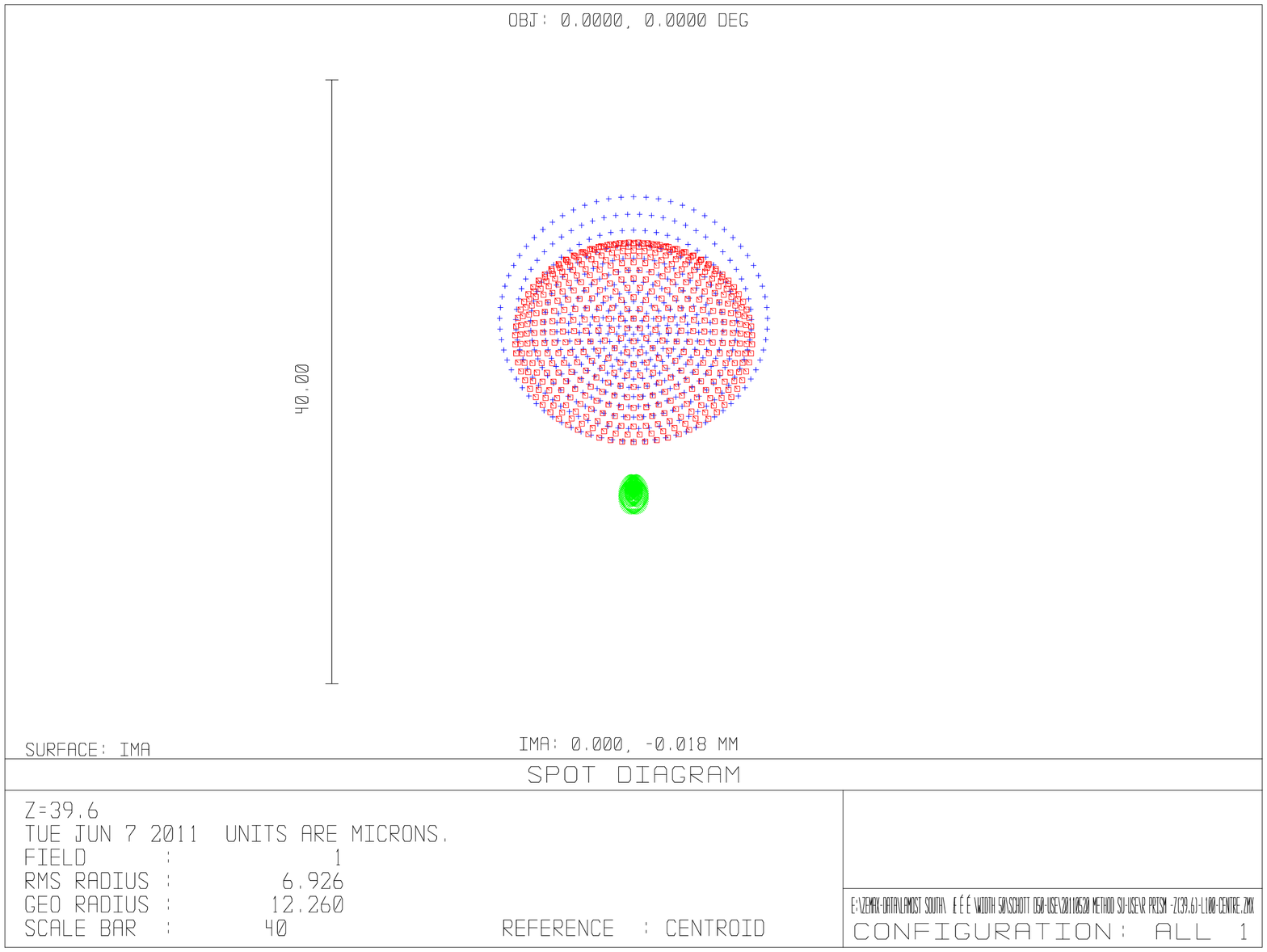}
  \includegraphics[scale=0.3,angle=90]{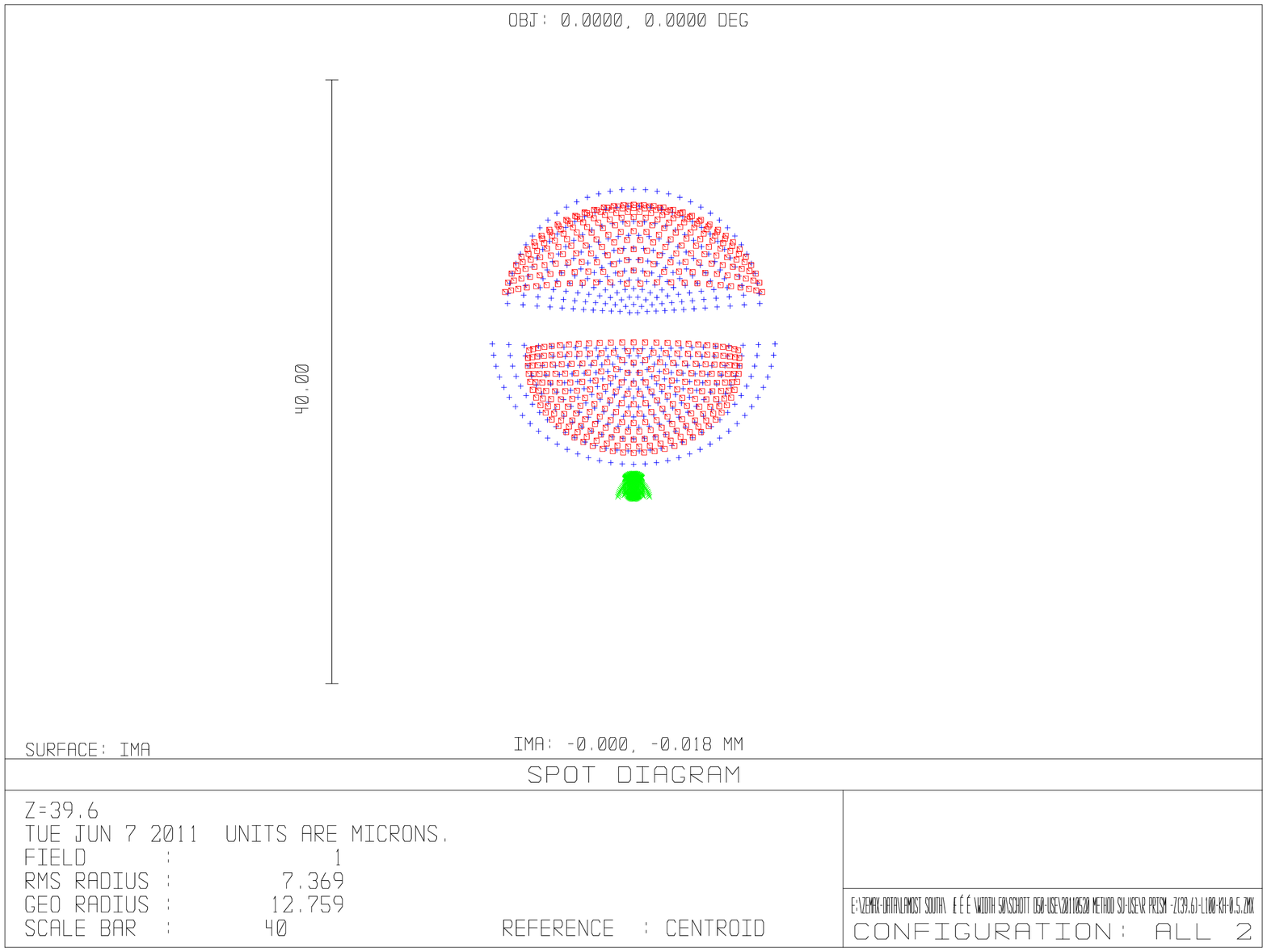}
  \caption{The spot diagram when the distance from the first surface of ADC to the focal surface is 100 $\mathrm{mm}$. The rest of explanation is the same as in Figure \ref{fig4}.}
  \label{fig7}
  \end{figure*}

\begin{figure*}
  \includegraphics[scale=0.3,angle=90]{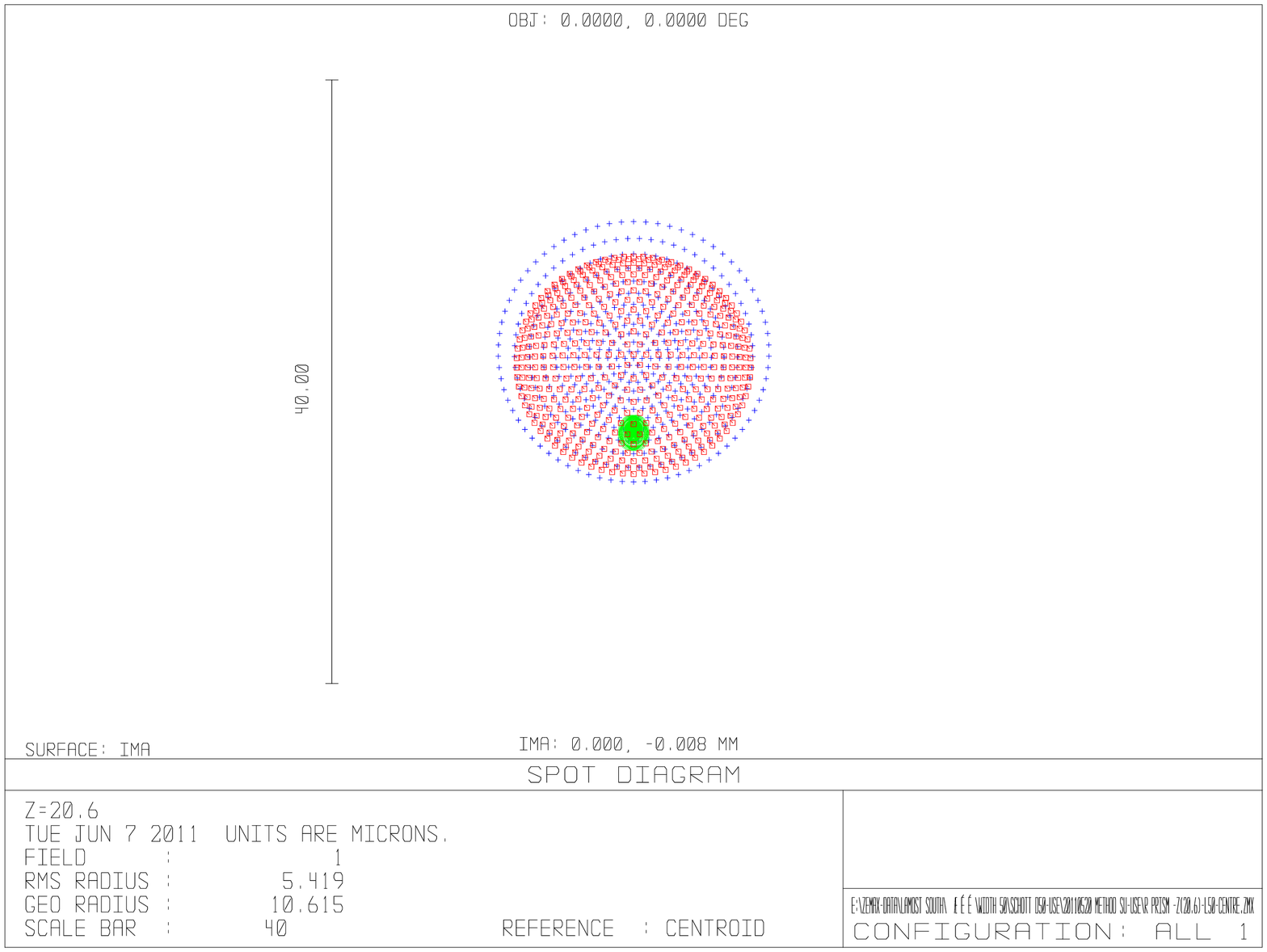}
  \includegraphics[scale=0.3,angle=90]{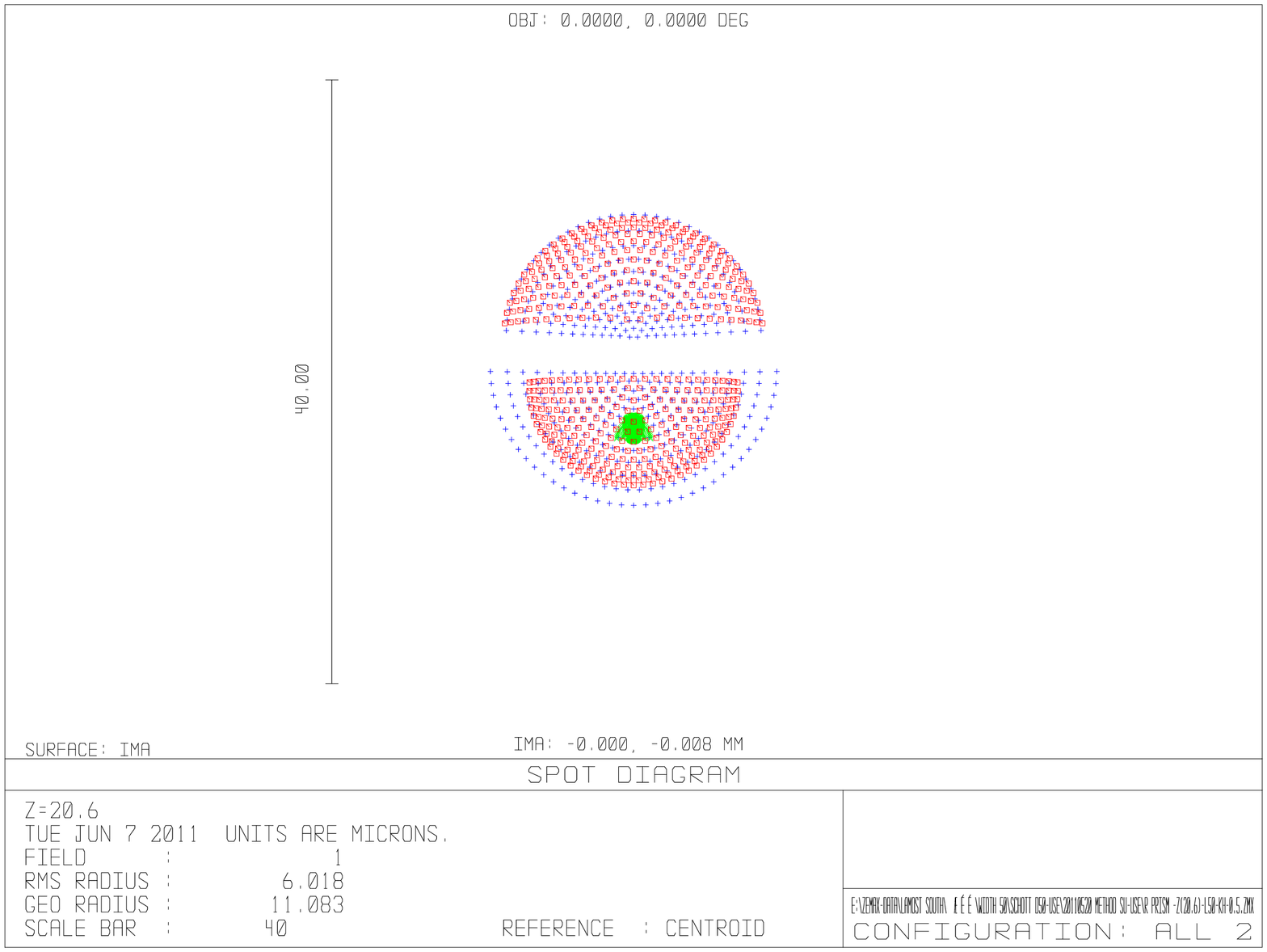}
  \caption{The spot diagram when the distance from the first surface of ADC to the focal surface is 50 $\mathrm{mm}$. The rest of explanation is the same as in Figure \ref{fig4}.}
  \label{fig8}
  \end{figure*}

The \textsc{Zemax} is used for the following calculations.\\
In this paper only the aberrations caused by the ADC are analyzed {\rmfamily i.e.} the aberrations of original optical system are ignored.\\
We take the two outside surfaces of the ADC to be concentric with the focal surface when this ADC is at the farthest position. In this situation if we ignore the structure of strip in the whole FOV the images are the same. We only need to calculate and discuss the case where the object is at the centre of FOV but with different relations to the strips. For centre objects we take two states: (1) the light beam area of object is at the middle of a lens--prism strip; (2) a slit of two lens--prism strips is at the centre of the light beam area of this object, {\rmfamily i.e.}, the light of this object is half in one strip and half in another strip. In this section all spot diagrams are calculated for these two states. For state (1) we adjust the tilt angle of the cemented surface of lens--prism strip to make the image centroids of $\lambda$ $=$ 380 $\mathrm{nm}$ and $\lambda$ $=$ 1000 $\mathrm{nm}$ coincide at $\mathrm{z}$ $=$ 65.4 $^\circ$, {\rmfamily i.e.}, to remove the atmospheric dispersion at these two extreme wavelengths. Thus the tilt angle 6.89 $^\circ$ mentioned above is obtained. In this situation we found at wavelength about 500 $\mathrm{nm}$, the image centroid is farthest from the co--centre of $\lambda$ $=$ 380 $\mathrm{nm}$ and $\lambda$ = $1000$ $\mathrm{nm}$, the angular distance of them is the maximum compensation error. Since all images should be in the one same focal surface in calculation of state (2) the focal surface position of state (1) is used.\\
Since the radii of ADC are fixed when the distance between ADC and the focal surface is less than 250 $\mathrm{mm}$, the two outside surfaces of ADC are not concentric with the focal surface. In this situation image quality in centre and off--axis of FOV are different. But the difference is small. We still only calculate and discuss where the object is at the FOV centre and in the above two states, {\rmfamily i.e.}, (1) the light beam area of object is at the middle of a lens--prism strip. For each decided position of ADC the zenith distance is chosen to make the image centroids of $\lambda$ = 380 $\mathrm{nm}$ and $\lambda$ $=$ 1000 $\mathrm{nm}$ coinciding, {\rmfamily i.e.}, eliminate the atmospheric dispersion at these two extreme wavelengths. (2) a slit of two lens--prism strips is at the centre of the light beam area of this object. In this state the focal surface position of state (1) is used.\\
These main calculation results are shown in Figure \ref{fig4}, \ref{fig5}, \ref{fig6}, \ref{fig7}, \ref{fig8} and Table \ref{tab3},\ref{tab4},\ref{tab5}. It is clear that the predictions in \ref{abe} and \ref{compen} are proved and some specific values are obtained: the largest monochromatic image at extreme wavelength is about 0.18 $\mathrm{arc sec}$, mainly is chromatic aberration. This figure shows no apparent change as the ADC moves along the optical axis. From Figure \ref{fig4}, \ref{fig5}, \ref{fig6}, \ref{fig7} and \ref{fig8} we can find some coma induced by tilt cemented surface and it reduces with the ADC's moves towards the focal surface. The maximum compensation error of atmospheric dispersion is 0.29 $\mathrm{arc sec}$ for the atmospheric dispersion 3.35 $\mathrm{arc sec}$, {\rmfamily i.e.}, the compensation error is about $1/12$ of the atmospheric dispersion. The total spread in the waveband between $\lambda$ $=$ 380 $\mathrm{nm}$ and $\lambda$ = 1000 $\mathrm{nm}$ is about 0.4 $\mathrm{arc sec}$. In Figure \ref{fig4}, \ref{fig5}, \ref{fig6}, \ref{fig7} and \ref{fig8}, the same wavelength images are separated about 0.04 $\mathrm{arc sec}$ because the light of this object is half in one strip and half in the other strip. These two tilt cemented surfaces have different distances to focal surface. They produce different dispersions, one bigger and the other smaller than the average dispersion. By the way, the most part of one semi--circle of lens--prism strip is LLF1, the other most part is PSK3, they produce the difference chromatic aberration, so the sizes of chromatic spread are different. Even though a 0.4 $\mathrm{arc sec}$ geometrical aberration is brought, it is worthy to use such a simple ADC to correct 3.35 $\mathrm{arc sec}$ atmospheric dispersion. Based on these analysis and calculation above, some extending results could be obtained. For example if the thickness of ADC increases to 18 $\mathrm{mm}$, the chromatic aberration increase about 0.085 $\mathrm{arc sec}$ and the total spread from waveband $\lambda$ $=$ 380 $\mathrm{nm}$ to $\lambda$ $=$ 1000 $\mathrm{nm}$ increases to about 0.45 $\mathrm{arc sec}$. In this situation the wide of each strip can increase to 75 $\mathrm{mm}$ or even 100 $\mathrm{mm}$, thus ADC at the farthest position only $2/3$ or even $1/2$ objects will meet a slit.\\

\section{Conclusion}
In this paper a new kind of atmospheric dispersion corrector (ADC) is put forward. It is a segmented lens which consists of many lens--prism strips. As an example, such an ADC is discussed and designed for the LAMOST South. Its linear diameter of FOV is 1.22 $\mathrm{m}$ and its f--ratio is 5. Although this ADC's diameter is 1.27 $\mathrm{m}$, its thickness is only 12 $\mathrm{mm}$. Thus the difficulty of obtaining big optical glass is avoided, and the aberration caused by the ADC is small. When we move this segmented lens along the optical axis, the different dispersions can be obtained. These slits of ADC will produce about 2.5 $\mathrm{per cent}$ average obstruction loss of light. In the largest zenith distance each celestial object's light only meets one slit of this ADC, and at other zenith distance only a part object light meet a slit. The effects of ADC slits on the diffraction energy distribution are discussed. Since LAMOST is used for optical fibre spectroscopic observation and the diameter of optical fibre is 1.6 $\mathrm{arc sec}$ (LAMOST South) and 3.3 $\mathrm{arc sec}$ (LAMOST North), these effects are acceptable. From waveband $\lambda$ $=$ 380 $\mathrm{nm}$ to $\lambda$ $=$ 1000 $\mathrm{nm}$ a total spread of aberration 0.4 $\mathrm{arc sec}$, which mainly is the compensation error of dispersions and achromatic aberration, will be brought when we use such an ADC for compensating 3.35 $\mathrm{arc sec}$ atmospheric dispersion. In this segmented ADC the diffraction spot is enlarged and the loss of light obstruction by slit is uneven. For these two reasons, such a segmented ADC is not suitable for the diffraction-limited high resolution telescope and photometry. It is mainly used for the optical fibre spectroscopic telescope. Since this ADC is thin, the aberration caused is small, there is no difficulty for obtaining its glasses, its thickness does not increase with the enlarging of FOV, and it has almost the same image quality for the center and off--axis of FOV. Such an ADC is especially suitable for very large FOV optical fibre spectroscopic telescopes.\\

\section*{Acknowledgments}

Thanks to Professor Xiangqun Cui for her enthusiastic support and helpful discussion, and to Professor Xiangyan Yuan for her helpful discussion and assistance during calculation.

\bibliographystyle{mn2e}
\bibliography{ADC}

\clearpage

    \clearpage

    \clearpage
    \begin{table*}
    \caption{\label{t1}The structure parameters of the ADC}
    \centering
    \begin{tabular}{ccccc}
    \hline\hline
    Surface&Radius ($\mathrm{mm}$)&Separation ($\mathrm{mm}$)&Glass&Tilt Angle ($^\circ$)\\
    1&20250& & & \\
     &     &6&LLF1& \\
    2&20244& & &6.89\\
     & &6&PSK3& \\
    3&20238& & & \\
     & &238& & \\
    Focal Surface&20000& & & \\
    \hline
    \end{tabular}
    \end{table*}

    \begin{table*}
    \centering
    \caption{\label{t2}The refractive index of the Schott glass PSK3 and LLF1}
    \centering
    \begin{tabular}{ccccccc}
    \hline\hline
    $\lambda$ ($\mathrm{nm}$) &380&441.8&500&587.6&656.3&1000\\
    LLF1&1.574977&1.562383&1.555066&1.548138&1.544564&1.535632\\
    PSK3&1.570682&1.562383&1.557313&1.552320&1.549650&1.542388\\
    \hline
    \end{tabular}
    \end{table*}

\clearpage

\clearpage

 \clearpage

  \clearpage

  \clearpage

\clearpage
\begin{table*}
\centering
\caption{\label{tab3}The distance from the first surface of ADC to the focal surface, corresponding to the object's zenith distance at Paranal Observatory and the diameter of the light beam area of the object on the ADC.}
\begin{tabular}{cccccc}
\hline\hline
Distance from the first surface \\of ADC to the focal surface ($\mathrm{mm}$)&250&200&150&100&50\\
& & & & & \\
The object's zenith distance\\at Paranal Observatory ($^\circ$)&65.4&59.9&51.9&39.6&20.6\\
 & & & & & \\
The diameter of the light beam area of \\the object on the ADC ($\mathrm{mm}$)&50&40&30&20&10\\
\hline
\end{tabular}
\end{table*}

\begin{table*}
\caption{\label{tab4} The correction results of the ADC when the light beam area of the object is at the middle of a lens--prism strip. Largest spread of spot diagram and compensation error are in $\mathrm{arc sec}$.}
\centering
\begin{tabular}{cccccc}
\hline\hline
Distance from the first surface \\of ADC to focal surface ($\mathrm{mm}$)&250&200&150&100&50\\
Largest spread of spot diagram $\lambda$ 380 $\mathrm{nm}$&0.180&0.180&0.182&0.183&0.185\\
Largest spread of spot diagram $\lambda$ 500 $\mathrm{nm}$&0.033&0.029&0.027&0.024&0.021\\
Largest spread of spot diagram $\lambda$ 1000 $\mathrm{nm}$&0.172&0.168&0.165&0.162&0.160\\
Largest spread of spot diagram \\of the whole waveband
$\lambda$ 380 $\mathrm{nm}$ -- 1000 $\mathrm{nm}$
&0.401&0.343&0.279&0.215&0.185\\
Compensation error&0.289&0.227&0.175&0.113&0.041\\
\hline
\end{tabular}
\end{table*}

\label{lastpage}
\begin{table*}
\caption{\label{tab5} The correction results of the ADC when a slit of two lens--prism strips is at the centre of the light beam area of the object. Largest spread of spot diagram and compensation error are in $\mathrm{arc sec}$.}
\centering
\begin{tabular}{cccccc}
\hline\hline
Distance from the first surface \\of ADC to focal surface ($\mathrm{mm}$)&250&200&150&100&50\\
Largest spread of spot diagram $\lambda$ 380 $\mathrm{nm}$&0.192&0.193&0.194&0.196&0.200\\
Largest spread of spot diagram $\lambda$ 500 $\mathrm{nm}$&0.028&0.025&0.023&0.026&0.027\\
Largest spread of spot diagram $\lambda$ 1000 $\mathrm{nm}$&0.188&0.187&0.185&0.183&0.184\\
Largest spread of spot diagram \\of the whole waveband
$\lambda$ 380 $\mathrm{nm}$ -- 1000 $\mathrm{nm}$
&0.397&0.339&0.276&0.212&0.200\\
Compensation error &0.289&0.227&0.175&0.108&0.046\\
\hline
\end{tabular}
\end{table*}
\clearpage
\end{document}